# New solutions of Heun's general equation

Artur Ishkhanyan[1] and Kalle-Antti Suominen[2,3]

[1]Engineering Center of Armenian National Academy of Sciences, Ashtarak-2, 378410 Armenia
[2]Helsinki Institute of Physics, PL 64, FIN-00014 Helsingin yliopisto, Finland
[3]Department of Physics, University of Turku, FIN-20014 Turun yliopisto, Finland

**Abstract.** We show that in four particular cases the derivative of the solution of Heun's general equation can be expressed in terms of a solution to another Heun's equation. Starting from this property, we use the Gauss hypergeometric functions to construct series solutions to Heun's equation for the mentioned cases. Each of the hypergeometric functions involved has correct singular behavior at only one of the singular points of the equation; the sum, however, has correct behavior.

Heun's equation is the most general linear Fuchsian equation of second order with four regular singularities [1], and it is therefore a natural generalization of the hypergeometric equation. Since the special cases of Heun's equation include the Gauss hypergeometric, confluent hypergeometric, Mathieu, Ince, Lame, Bessel, Legendre, Laguerre equations, etc, it is of considerable importance in mathematical physics.

The canonical form of Heun's general equation, whose Klein–Bŏcher–Ince formula [2] is [0, 4, 0], is given as

$$u_{zz} + \left(\frac{\gamma}{z} + \frac{\delta}{z-1} + \frac{\varepsilon}{z-a}\right)u_z + \frac{\alpha\beta z - q}{z(z-1)(z-a)}u = 0, \qquad (1)$$

where the subscripts stand for differentiation with respect to the corresponding variable and the parameters satisfy the Fuchsian condition $1 + \alpha + \beta = \gamma + \delta + \varepsilon$. The singular points of this equation are located at $z = 0, 1, a$ and $\infty$. The solution to this equation is denoted as $u = H(a, q; \alpha, \beta, \gamma, \delta; z)$ (throughout this paper we omit the non-essential constant multiplier).

By taking the derivative of (1) and then subtracting from the obtained equation the initial equation (1) multiplied by some function $\psi(z)$, we obtain a third-order linear equation of the form

$$u_{zzz} + (f - \psi)u_{zz} + (f_z + g - \psi f)u_z + (g_z - g\psi)u = 0, \qquad (2)$$

where $f$ and $g$ denote the coefficients of equation (1) at $u_z$ and $u$, respectively. Obviously, this equation becomes a second-order equation for $u_z$ when $g_z - g\psi = 0$:

$$u_{zzz} + f_1 u_{zz} + g_1 u_z = 0, \tag{3}$$

the coefficients $f_1$ and $g_1$ then being explicitly given as

$$f_1 = \left(\frac{\gamma+1}{z} + \frac{\delta+1}{z-1} + \frac{\varepsilon+1}{z-a}\right) - \frac{\alpha\beta}{\alpha\beta z - q}, \tag{4}$$

$$g_1 = -\left(\frac{\gamma}{z^2} + \frac{\delta}{(z-1)^2} + \frac{\varepsilon}{(z-a)^2}\right) + \frac{\alpha\beta z - q}{z(z-1)(z-a)} + \left(\frac{\gamma}{z} + \frac{\delta}{z-1} + \frac{\varepsilon}{z-a}\right) \\ \times \left[\left(\frac{1}{z} + \frac{1}{z-1} + \frac{1}{z-a}\right) - \frac{\alpha\beta}{\alpha\beta z - q}\right]. \tag{5}$$

As can be seen, equation (3), in general, has five singular points: we have one more singularity at the point $z = q/\alpha\beta$. However, evidently, in four particular cases, when this point coincides with already existing singularities, i.e., when $q = 0$, $q = \alpha\beta$, $q = a\alpha\beta$ and $\alpha\beta = 0$, the number of singularities remains four. Since the singularities remain regular, the derived equation can be transformed to Heun's general equation. Then the conclusion is that **_in the four listed cases the derivative of the solution to the initial Heun's equation (1) can be expressed in terms of a solution to another Heun's equation_**. As will be demonstrated below, this observation could be useful since it may lead to interesting series solutions to Heun's equation in terms of hypergeometric functions.

After some algebra, we arrive at the following final formulas.

1. $q = 0$.

$$\frac{d}{dz} H(a,0;\alpha,\beta,\gamma,\delta;z) = z^s H(a,q';\alpha',\beta',2s+\gamma,\delta+1;z), \quad s = 1 \text{ or } -\gamma, \tag{6}$$

$$\alpha'\beta' = \alpha\beta + (2\gamma + \delta + \varepsilon) + s(\delta + \varepsilon + 2), \tag{7}$$

$$1 + \alpha' + \beta' = \gamma + \delta + \varepsilon + 2(s+1), \tag{8}$$

$$q' = \gamma(1+a) + s[a(\delta+1) + (\varepsilon+1)]. \tag{9}$$

2. $q = \alpha\beta$.

$$\frac{d}{dz} H(a,\alpha\beta;\alpha,\beta,\gamma,\delta;z) = (z-1)^s H(a,q';\alpha',\beta',\gamma+1,2s+\delta;z), \quad s = 1 \text{ or } -\delta. \tag{10}$$

$$\alpha'\beta' = \alpha\beta + (\gamma + 2\delta + \varepsilon) + s(\gamma + \varepsilon + 2), \tag{11}$$

$$1 + \alpha' + \beta' = \gamma + \delta + \varepsilon + 2(s+1), \tag{12}$$

$$q' = q + (\gamma + \varepsilon + a\delta) + a(\gamma+1)s. \tag{13}$$

3. $q = a\alpha\beta$.

$$\frac{d}{dz}H(a,a\alpha\beta;\alpha,\beta,\gamma,\delta;z) = (z-a)^s H(a,q';\alpha',\beta',\gamma+1,\delta+1;z), \quad s=1 \text{ or } -\varepsilon. \tag{14}$$

$$\alpha'\beta' = \alpha\beta + (\gamma+\delta+2\varepsilon) + s(\gamma+\delta+2), \tag{15}$$

$$1+\alpha'+\beta' = \gamma+\delta+\varepsilon+2(s+1), \tag{16}$$

$$q' = a\alpha\beta + a(\gamma+\delta) + \varepsilon + s(\gamma+1). \tag{17}$$

4. $\alpha\beta = 0$.

$$\frac{d}{dz}H(a,q;0,\beta,\gamma,\delta;z) = H(a,q';\alpha',\beta',\gamma+1,\delta+1;z), \tag{18}$$

$$\alpha'\beta' = 2(\gamma+\delta+\varepsilon), \tag{19}$$

$$1+\alpha'+\beta' = (\gamma+\delta+\varepsilon)+3, \tag{20}$$

$$q' = q + \gamma(1+a) + \delta a + \varepsilon. \tag{21}$$

The above relations are not trivial. This can be understood, for instance, if we note that they lead to closed form solutions to the general Heun's equation (1) for some particular sets of parameters. As an illustrative example, consider the first case, $q=0$. If we additionally demand $\varepsilon = -1$ and $q' = a\alpha'\beta'$, we see that the Heun function on the right-hand side of (6) becomes a Gauss hypergeometric function, so that equation (6) can be resolved in terms of hypergeometric functions. If, for instance, $s=1$, (6) is reduced to

$$\frac{d}{dz}H(a,0;\alpha,\beta,\gamma,\delta;z) = z \cdot {}_2F_1(\alpha',\beta';\gamma+2;z), \tag{22}$$

where the parameters of the hypergeometric function are to be defined from relations (7) and (8) ($\varepsilon = -1$) and additionally should be

$$\gamma + \alpha = a(\alpha\beta + \gamma + \delta). \tag{23}$$

Equation (22) is then easily integrated to give a solution of Heun's equation involving a combination of two Gauss hypergeometric functions. This solution is a particular case of the solution by Craster [3].

Now we will show that the above relations can be used for construction of series solutions to the Heun equation in terms of hypergeometric functions.

Consider, for example, the case $q = a\alpha\beta$. Denoting Heun's function on the right-hand side of equation (14) by $v$, then applying further transformation

$$v = z^{-\gamma}(z-1)^{-\delta} w \tag{24}$$

equation (14) is rewritten in the following form:

$$\frac{d}{dz}H(a,a\alpha\beta;\alpha,\beta,\gamma,\delta;z)=z^{-\gamma}(z-1)^{-\delta}(z-a)^{s}H(a,q_{1};\alpha_{1},\beta_{1},1-\gamma,1-\delta;z), \quad s=1\text{ or }-\varepsilon, \quad (25)$$

$$\alpha_{1}\beta_{1}=\alpha\beta+(\varepsilon+s)(2-\gamma-\delta), \quad (26)$$

$$1+\alpha_{1}+\beta_{1}=2-(\gamma+\delta+\varepsilon)+2(\varepsilon+s), \quad (27)$$

$$q_{1}=a\alpha\beta+(\varepsilon+s)(1-\gamma). \quad (28)$$

Now putting $s=-\varepsilon$ ($\Rightarrow \alpha_{1}=-\alpha, \beta_{1}=-\beta, q_{1}=a\alpha\beta$) and using the power series expansion of Heun's function on the right-hand side of equation (25):

$$H(a,a\alpha\beta;-\alpha,-\beta,1-\gamma,1-\delta;z)=\sum_{n=0}^{+\infty}a_{n}z^{n} \quad (29)$$

with $a_n$ obeying a three term recursion relation (see [1]), we resolve (25) to get

$$H(a,a\alpha\beta;\alpha,\beta,\gamma,\delta;z)=\sum_{n=0}^{+\infty}a_{n}\left(\int z^{n-\gamma}(z-1)^{-\delta}(z-a)^{-\varepsilon}dz\right). \quad (30)$$

In general, the integrals involved in this sum are the Appell hypergeometric functions of two variables of the first kind [4]:

$$\int z^{n-\gamma}(z-1)^{-\delta}(z-a)^{-\varepsilon}dz = \frac{(-1)^{-\delta-\varepsilon}}{a^{\varepsilon}}\frac{z^{1+n-\gamma}}{1+n-\gamma}F_{1}\left(1+n-\gamma;\delta,\varepsilon;2+n-\gamma;z,\frac{z}{a}\right), \quad (31)$$

$$F_{1}(a;b_{1},b_{2};c;x,y)=\sum_{m,n=0}^{+\infty}\frac{(a)_{m+n}(b_{1})_{m}(b_{2})_{n}}{(c)_{m+n}}\frac{x^{m}}{m!}\frac{y^{n}}{n!}. \quad (32)$$

However, in several cases they reduce to simpler functions. For example, obviously, if $\varepsilon$ or $\delta$ are negative integers, the Appell function is expressed through a finite number of Gauss hypergeometric functions. Another example is the case $\varepsilon=\delta$ and $a=-1$, when we have one hypergeometric function of the argument $z^2$:

$$\int z^{n-\gamma}(z^{2}-1)^{-\delta}dz=\frac{(-1)^{-\delta}z^{1+n-\gamma}}{1+n-\gamma}{}_{2}F_{1}\left(\frac{1-\gamma+n}{2},\delta;\frac{3-\gamma+n}{2};z^{2}\right). \quad (33)$$

A different possibility appears when we put $s=1$ in equation (25). Again using the power series expansion of Heun's function on the right-hand side with parameters given by (26)-(28) at $s=1$ ($\gamma_{1}=1-\gamma, \delta_{1}=1-\delta, \varepsilon_{1}=2+\varepsilon$):

$$a_{n}an[n-1+\gamma_{1}]-a_{n-1}[(1+a)(n-1)(n-2)+(\gamma_{1}(1+a)+a\delta_{1}+\varepsilon_{1})(n-1)+q_{1}]$$
$$+a_{n-2}[(n-2)(n-3)+(\gamma_{1}+\delta_{1}+\varepsilon_{1})(n-2)+\alpha_{1}\beta_{1}]=0, \quad n\geq 0, \quad (34)$$

$$a_{-2}=a_{-1}=0, \quad a_{0}=1, \quad a_{1}=q_{1}/a\gamma_{1},$$

we arrive at the following expansion in terms of incomplete beta functions:

$$H(a,a\alpha\beta;\alpha,\beta,\gamma,\delta;z)=\sum_{n=0}^{+\infty}a_{n}[aB_{z}(1-\gamma+n,1-\delta)-B_{z}(2-\gamma+n,1-\delta)]. \quad (35)$$

This series can be rewritten in terms of Gauss hypergeometric functions

$$H(a,a\alpha\beta;\alpha,\beta,\gamma,\delta;z) = z^{1-\gamma}\left[\frac{a}{1-\gamma}\cdot{}_2F_1(1-\gamma,\delta;2-\gamma;z)\right. \qquad (36)$$
$$\left. + (a-1)\sum_{n=1}^{+\infty} a_n \frac{z^n}{1-\gamma+n}\cdot{}_2F_1(1-\gamma+n,\delta;2-\gamma+n;z)\right].$$

Finally, as can be easily shown, similar expansions can be constructed for the other three cases, when $q=0$, $q=\alpha\beta$ and $\alpha\beta=0$.

There have been many studies towards solving Heun's equation by means of a series of hypergeometric functions (for a summary see [1]). However, the chosen forms of the hypergeometric functions are different from those used here. Indeed, in the earlier papers by Svartholm, Erdelyi and Schmidt [5], the Gauss hypergeometric functions were used in the summations that have correct singular behaviour at *two* fixed singular points, and then the summation was used in attempting to correct for that at the other points. A non-trivial different approach was recently suggested by Craster [3]. The idea was to use a combination of two (at $\varepsilon=-1$, and $N+1$ in the more general case $\varepsilon=-N$) hypergeometric functions, again having correct singular behavior at two singular points, each of which has slightly incorrect singular behavior at *a different* singular point. The idea here qualitatively differs from the approaches used by above authors. We use hypergeometric functions having correct singular behavior only at *one* singular point, *the same for all the involved hypergeometric functions*, in order to construct a sum that, however, has the correct behavior. This is immediately seen from equation (36), since the functions involved in the expansion are given by the following Riemann $P$-function:

$$P\begin{pmatrix} 0 & 1 & \infty \\ 0 & 0 & 1-\gamma+n \\ -1+\gamma-n & 1-\delta & \delta \end{pmatrix}. \qquad (37)$$

Furthermore, if compared with the approach by Craster *et al*, we operate with infinite series instead of the finite sums used in [3]. And the solutions apply to different region of the parameters: in general, we impose only one restriction on the parameters of Heun's equation, while Craster's finite sums always demand two of them. Finally, the conditions for termination of our series do not coincide, in general, neither with those considered by Craster *et al* nor with those of [5]. It is well understood that our results could not be deduced from the approaches used by previous authors.

**Acknowledgments**

The authors acknowledge support from the Academy of Finland (Project 50314) and from the Armenian National Science and Education Fund (grant no. PS-10). AI thanks the Helsinki Institute of Physics for kind hospitality and Ashot Manukyan for useful discussions, as well as acknowledging the support by the International Science and Technology Center (grant A215) and US Civilian Research and Development Foundation (grant no NFSAT PH 100-02/12042).